# A Set-Theoretic Metaphysics for Quantum Mechanics

Paul Tappenden  paulpagetappenden@gmail.com

24th September 2023

The lesson to be learned from what I have told of the origin of quantum mechanics is that probable refinements of mathematical methods will not suffice to produce a satisfactory theory, but that somewhere in our doctrine is hidden a concept, unjustified by experience, which we must eliminate to open up the road.

(Born 1954: 11)

May the spirit of Newton's method give us the power to restore unison between physical reality and the profoundest characteristic of Newton's teaching – strict causality.

(Einstein 1927: 467)

## ABSTRACT

Set theory brought revolution to philosophy of mathematics and it can bring revolution to philosophy of physics too. All that stands in the way is the intuition that sets of physical objects cannot themselves be physical objects, which appears to depend on the ubiquitous assumption that it's possible for there to exist numerically distinct observers in qualitatively identical mental states. Overturning that assumption opens the way to construing an object in superposition in an observer's environment as a set of objects in definite states. The components of the superposition are subsets for which all the elements are in the same definite state. So an *environmental* z-spin-up electron becomes a set of *elemental* electrons each of which has definite spin for one orientation but lacks indefinite spin for other orientations. The *environmental* z-spin-up electron has subsets of *elemental* electrons for every orientation but it's only the subset with spins on the z-axis for which all the elements of the subset have the same value, namely spin-up. The subset of *elemental* electrons with spins on the x-axis has subsets of spin-up and spin-down *elemental* electrons of equal measure. Observers only detect the spins of *environmental* electrons, not those of *elemental* electrons.

## 1 Hidden Mechanics

On the 24th September 1923 Louis de Broglie made this momentous prediction:

> Le nouveau principe mis à base de la dynamique expliquerait la diffraction des atomes de lumière, *si petit que soit leur nombre*. De plus un mobile quelconque pourrait dans certains cas se diffracter. Un flot d'électrons traversant une ouverture assez petite présenterait des phénomènes de diffraction. C'est de ce côté qu'il faudra peut-être chercher des confirmations expérimentales de nos idées.[1]
>
> (de Broglie 1923: 549, original emphasis)

Notice that he emphasised that he was predicting that *single* electrons would diffract. How can the world *be* like that? A century later, there's no consensus on how, and much effort is being made to establish that physics can do without the concept of a mind-independent "external" world altogether (Müller 2020). Contrariwise, effort continues to be put into preserving realism. What follows is in that vein and attempts to succinctly present ideas which have been developed in a series of papers; clarifying, correcting and adding further thoughts (Tappenden 2017, 2021, 2022, 2023).

I'll argue that what we should eliminate "to open up the road" is the assumption that sets of physical objects are not themselves physical objects. An *environmental* double-slit apparatus could be a set of *elemental* apparatuses which are macroscopically isomorphic but microscopically anisomorphic, in each of which a single *elemental* particle passes through a single *elemental* slit. "Elemental" in the set-theoretic sense. The dichotomy between an *environmental* realm and an *elemental* realm will be fundamental, so I'll italicise those terms throughout. A single electron in an observer's environment can be a set of *elemental* electrons, each following a different trajectory in an *elemental* universe.

This is Many Worlds with hidden-variables, like Many Interacting Worlds theory (Hall *et al.* 2014; Sebens 2015; Boström 2015). The difference is that observers inhabit sets of universes rather than individual universes, with the result that the theory provides further insight into the issues of causal locality and the separability of entangled particles.

---

[1] This new basic principle of dynamics would explain the diffraction of particles of light, *however small their number*. Further, any moving object could diffract in the right circumstances. A stream of electrons passing through a sufficiently small opening would display diffraction phenomena. It's here that we should perhaps look for experimental confirmation of our ideas. (author's translation)

The application of set theory to the metaphysics of quantum mechanics resolves the longstanding ambivalence about the ontic status of wavefunction for Many Worlds which is clearly illustrated in Sean Carroll's recent book when he writes:

> What the World Is Made Of… A quantum wave function
> (Carroll 2019: 49)
>
> wave functions are superpositions of different possibilities
> (*ibid.*: 64)
>
> Wave functions may be real, but they're undeniably abstract
> (*ibid.*: 79)

Concrete objects are undeniably abstract, and made of possibilities? The ambivalence arises from following Hugh Everett III in thinking of Many Worlds as being a *pure wave* theory (Everett 1957). Impurity is required in the form of hidden-variables taking definite values, but if the difficulties of extant hidden-variable theories are to be avoided the definite values must not attach to objects which exist in observers' environments. However, they can harmlessly exist in the set-theoretic elements of observers' environments.

If the double-slit apparatus is a set of apparatuses it follows that the observer's body is a set of bodies. The observer *splits* Everett-style into observers making different electron-detection observations because their body partitions into cerebrally anisomorphic bodies which are subsets instancing those different observations. So, eliminating the idea that sets of physical objects are not themselves physical objects requires eliminating the idea that there's a one-to-one relation between observers and bodies. The core conceptual change required is to drop the ubiquitous assumption enshrined in a seminal thought experiment by Hilary Putnam which has had a profound influence on contemporary analytic philosophy of mind (Putnam 1975). So, the next section is on metaphysics of mind, to prepare the way for metaphysics of physics, followed by a conclusion and its consequences.

## 2 Mind in a Multiverse

The set-theoretic metaphysics involves an infinite set of universes, a muiltiverse. They are to be thought of as distributed in configuration space. The difference with classical statistical mechanics is that configurations constitute a space of actualities rather than an imaginary space of possibilities.

Fundamental to making sense of this is our conception of how observers are *situated* in a multiverse, to use Jeffrey Barrett's term (Barrett 2021). The metaphysics has to be consistent with observed phenomena and phenomena depend on how observers are situated. A useful step in thinking about how an observer is situated in a multiverse is to temporarily set aside quantum theory and consider what Max Tegmark has called the Level I multiverse (Tegmark 2007).

According to current cosmology there's no evidence that space is finite. The *observable universe* surrounds us out to a distance of about 46 billion lightyears. Beyond that, we have no reason to believe that the overall pan-galactic structure changes from place to place, so there can be duplicate observable universes. Tegmark has estimated that the average separation of universes observationally isomorphic to ours is $10^{10^{117}}$ metres. So far as we know, our observable universe may be one of a denumerably infinite number of copies.

There's a metaphysical bias in that last sentence. Is *our* observable universe one among many, or is our universe the very collective itself? If the collective, it can't be the mereological sum because that would entail that our observable universe has infinite mass which isn't the case. However, our observable universe could be the *set* of individual universes so long as we relinquish the idea that sets are necessarily abstract objects and adopt this hypothesis:

Concrete Sets

A set of physical objects has all and only the properties which its elements share, other than those that are logically excluded.

Note that this does not imply that the *unit set* of a set of physical objects is itself a physical object; we can continue to think of that set as being abstract, except in the case of physical objects which are self-membered singletons, known to logicians as *Quine atoms*; see (Tappenden 2022: §4).

Concrete Sets has surreal consequences. For instance, the set of a sapphire and an emerald is a third gem which has definite mass if its elements have the same mass, together with *indefinite* colour and position, since its elements don't have the same colour and position. How could that possibly be the case? Where would the extra mass *be*? A thought experiment helps.

Consider two isomorphic observable universes similar to our own. We see a pair of doppelgängers beside tables on which are matched boxes. The boxes are special in that their interiors are causally screened from the surrounding environment; they could fittingly be called Schrödinger boxes. One contains the sapphire, the other the emerald.

The doppelgängers both emit noises which sound like the utterance "I see one box". Conventionally, the noises would be interpreted as tokens of two utterances by two distinct observers, each referring to one of the boxes, but an alternative interpretation is possible, given Concrete Sets.

There's a single observer who makes a single utterance tokened by the *set* of vocal noises. That observer refers to the box which is the set of the boxes containing the sapphire and the emerald. The single observer's box contains a *sappherald* a gem with indefinite colour and position but nonetheless a massive object in that observer's environment. The missing mass exists in the environment of the missing observer, so to speak, the *single* observer who's missing when the observer-doppelgänger relation is assumed to be one-to-one, as in Putnam's *Twin Earth* thought experiment. For more on the consequences of re-interpreting Twin Earth in this way see (Tappenden 2021: §5)

When the single observer opens their box the pair of doppelgänger make parallel movements in the two universes. As the box opens the doppelgänger are exposed to different visual stimuli and so emit different noises, one sounding like "Ah, a sapphire!" and the other sounding like "Ah, an emerald!" We can interpret what's happened as the original observer having split, Everett-style, into two observers making different observations. This suggests a conceptual connection between set theory, Everettian fission and the existence of objects with indefinite properties in quantum theory.

That's a third-person alternative analysis of how observers are situated in a multiverse inhabited by doppelgängers, but how does that look from the first-person point of view, assuming that a denumerably infinite number of copies of our observable universe do indeed exist? Isn't the concept of space itself somehow undermined?

No. All that's changed is our conception of how the universe beyond our observable universe is constituted. From the *Unitary Mind* point of view, beyond the observable universe is a massive macroscopic "superposition" of every way things have evolved since the Big Bang. That must be so because what we refer to as our observable universe is to be thought of as a set of universes which are observationally identical but surrounded by different distributions of matter and energy. Even without quantum theory, the concept of a spatially infinite universe coupled with Unitary Mind brings a fundamental dichotomy to material existence *relative to observers*. *Environmental* objects are sets of *elemental* objects.

I should mention that Nick Bostrom has briefly considered and rejected Unitary Mind, which he calls *Unification* (Bostrom 2006: 186). He endorses the usual intuition about the mind-body relation when he writes:

> It would, to say the least, be odd to suppose that whether one's own brain produces phenomenal experience strongly depends on the happenings in other brains that may exist in faraway galaxies that are causally disconnected from our solar system
>
> (*ibid*.: 188)

This doesn't take into account that the alternative interpretation of the mind-body relation entails an alternative interpretation of the constitution of *environmental* objects. According to the set-theoretic metaphysics the *elements* of an observer's environment don't *need* to be causally connected, they could, from a god's eye view, be multiple observable universes scattered through space.

When it comes to adding quantum theory to a Level I multiverse, think about the decay of an unstable particle and suppose, for the sake of argument, that the process is *stochastic*. An *environmental* particle is an infinite set of *elemental* particles, each of which has a *propensity* to decay given by its half-life. *Necessarily*, since the set is infinite, after the half-life has elapsed the original set of particles has partitioned into decayed and undecayed subsets of equal measure. An observer measuring the *environmental* particle after one half-life would fission into observers seeing and not seeing decay, and the subset measures of the downstream observers' bodies would necessarily be equal because that's the *probability measure*. The observer fissions because their body partitions; the particle fissions because *it* partitions. In fact the particle will be constantly partitioning into undecayed and decayed subsets, with the measure of the former decreasing and that of the later increasing until the measures are equal after one half-life.

So, we have it that according to Unitary Mind an observer who is well-informed about their situation in a stochastic Level I multiverse, and who is about to observe an unstable particle one half-life after its creation, should expect to fission into two observers making different observations in two distinct environments whose probability measures relative to the original environment are 0.5. What should the original observer *expect to observe*? This raises questions about persistence and probabilistic expectation which need to be resolved. Once that's done, we can go on to replace the concept of stochastic quantum processes with that of *dendritic* processes, as suggested by Everett (*op.cit.*: 460).

2.1 Fission and Persistence

Independently of quantum theory, the problem posed by fission as applied to persons was extensively discussed by Derek Parfit, who concluded that the concept of continuing identity must be abandoned in such situations (Parfit 1984). In that case, our observer who expects to fission has no warrant for expecting that *they* will observe anything at all! The problem is straightforward and applies to objects in general, not just persons. If X fissions into Y and Z and Y is not Z then Y *and* Z cannot *both* be X. So, if there's no reason to believe that Y *or* Z is X there's no reason to believe that X will persist beyond fission.

A way out of this impasse was introduced by Ted Sider and has come to be known as *stage theory* (Sider 1996; Hawley 2001: Ch.2). It was first explicitly applied to Everettian theory in (Tappenden 2008). The idea is that what a persisting thing *is* at time *t* is a temporal part (stage) of its history. Earlier and later stages of the history are the past and future *temporal counterparts* of the object at time *t* to which that object bears the relations *was* and *will be*. So X, Y and Z are distinct objects; X will be Y and X will be Z. For details of how X is related to the composite Y+Z see (Tappenden 2022: §2.1).

Similarly, Y was X and Z was X. An object persists by virtue of having past and future temporal counterparts. If it lacks past temporal counterparts then it was not an object in the past and if it lacks future temporal counterparts it will not be an object in the future.

Our single observer, knowingly about to fission, can expect to be two different observers making different observations in different environments, each with probability measure 0.5. How is the observer to make sense of that? Only via the concept of probabilistic expectation, which requires that the pre-fission observer should be uncertain about their future. But how can they be uncertain about their future if certain that they will fission?

2.2 Fission and Uncertainty

We're considering Unitary Mind in a Level I stochastic multiverse. The concept of stochasticity takes with it the concept of *objective* probability. If the decay of an *elemental* unstable particle is stochastic then it's an objective mind-independent property of that particle that it's probability of decay after one half-life is 0.5. So, since an observer's environment is an infinite set of *elemental* environments, the objective probability of decay of an *environmental* unstable particle for one half-life *is* the measure of the decayed subset after one half-life: 0.5.

Certainty and uncertainty are subjective states of observers. Assuming what's become known as the Principal Principle, an observer who knows that the objective probability of a future event is *p* should assign a subjective probability, a degree of belief or *credence* equal to *p*. For more on the Principal Principle in this context see (Tappenden 2021: §2.2). Our observer about to observe an *environmental* unstable particle after one half-life should thus assign a credence of 0.5 to observing decay and a credence of 0.5 to not observing decay. Our observer is uncertain about what they will observe *in exactly the same way* as an observer in a single stochastic universe.

This can seem odd because our observer must be, at the same time, *certain* that the *environmental* particle will be a particle which has decayed and will be a particle which has not decayed since the objective probability for that outcome is the sum of the measures of the undecayed and decayed subsets. But there's no contradiction here. The fact that both outcomes will occur entails that each will occur *even though the objective probabilities of each outcome are 0.5*. Our observer is uncertain as to what they will *observe* whilst certain as to what will *occur*. Our observer is certain that the *environmental* particle will be a decayed *environmental* particle even thought the objective probability of decay is 0.5. That's just a logical consequence of the fact that the environmental particle will be an undecayed *environmental* particle *and* will be a decayed *environmental* particle.

2.3 Finite versus Infinite[2]

All this plays havoc with our usual intuitions about probabilistic expectation because we're used to thinking in terms of *alternative possible* futures not *coexistent actual* futures. Intuition is further put to the test by thinking about stage theory and Unitatry Mind in the context of an adaptation of what Parfit called the *The Branch Line Case* (*op.cit.*: 287). He imagined a Star Trek style teleporter involving a transmitter on Earth and a receiver on Mars. On pressing the *send* button in the transmitter the prospective traveller is anaesthetised, their body scanned and all the necessary information is sent to the receiver where a copy is made and the original body destroyed. In the Branch Line Case there's a malfunction so that the original body isn't destroyed.

---

[2] Special thanks to Devin Bayer for raising this issue.

Imagine two isomorphic cubicles here on Earth. You are to go into one of them where you'll be anaesthetised and a copy of your body made in the other. The two doppelgängers will then awake and shortly afterwards a blue light will shine in the cubicle you went into and a green light in the other. Knowing all this in advance, what should you expect?

According to Unitary Mind you should expect to be an observer who wakes in the third cubicle which is the set of the two. That observer will then fission into one who sees blue and one who sees green. A first thought might be that you should assign equal credences to each future observation, but that's not warranted by the subset measures for the blue-lit cubicle and the green-lit cubicle. Although those subset measures are both 0.5 they are just measures on a set of two cubicles so they're not probability measures. Probability measures can only exist as measures on infinite sets, hence the longstanding problem of explaining probabilities in terms of frequencies.

However, another thought could seem to justify the equal credences, inspired by Lev Vaidman's introduction of the concept of post-measurement, pre-observation uncertainty (Vaidman 1998: 253). Suppose that the doppegängers were blindfolded when the lights went on and a bell rung in one cubicle and a gong struck in the other, but you're not told in advance which sound goes with which colour. You can expect to be an observer under a blue light and an observer under a green light, neither knowing which light they're under. So you can expect to be in a state of *self-location uncertainty,* each future observer assigning a credence of 0.5 to seeing a particular colour because they know that they're one of two observers who will see different colours when the blindfolds are removed. Bearing that in mind, it can seem reasonable to assigning equal credences to the future seeing of blue and the future seeing of green without intermediate blindfold, bell and gong.

Now imagine that there are *three* matched cubicles; you go into one, which will be blue-lit and copies of your body will be made in the other two, which will both be green-lit. Repeat the blindfold scenario, with a bell for blue and a gong for green and, again, you will be an observer under a blue light and you will be a *single* observer under a green light each not knowing which they are. So self-location uncertainty again yields equal credences for the future seeing of blue and the future seeing of green. However many green-lit cubicles there are, it *makes no difference* to your expectation of future experience based on future self-location uncertainty. And however many green-lit cubicles there are the subset measures of blue-lit and green-lit cubicles are *never* probability measures because the imaginary scenario can only ever involved a *finite* number of cubicles.

The predicament of observers in a Level I stochastic multiverse, given Unitary Mind and stage theory, is fundamentally different from that of observers in any Parfittian setup. But stochasticity is not for us. In the spirit of Albert Einstein's sentiment in the opening quote, Everett sought to replace the concept of stochastic process with that of a *deterministic* dendritic process. He wrote:

> The theory based on pure wave mechanics is a conceptually simple, *causal* theory
>
> (*op.cit.*: 462, my emphasis)

Everett had a truly revolutionary idea, but there's still no agreement amongst Many Worlds theorists about what a fully satisfactory version of the theory is. Everett put the cat amongst the pigeons of everyday intuition. What's being argued here is that the theory should be *both* deterministic *and* objectively probabilistic. The absolute square of amplitude of the Everettian *branches* which issue from dendritic quantum processes just *is* objective probability. Further, we need to abandon Everett's idea that the theory should involve a *pure wave* mechanics.

David Wallace is proposing a defence of pure wave theory involving an object-less metaphysics for quantum mechanics which appeals to Ontic Structural Realism (Wallace 2022). The project is ongoing and I wish it well; the challenge of reconciling quantum mechanics and general relativity is waiting in the wings. However, what follows demonstrates that quantum mechanics on its own can be made fully intelligible with an ontology of objects bearing definite physical properties. Again, it becomes analogous to statistical mechanics, the difference being that configuration space becomes a space of actualities rather than possibilities and the interactions of objects in space is explained in terms of interactions between their set-theoretic elements in configuration space. A Humean question remains: how is natural necessitation possible? If making sense of quantum mechanics takes us back to a clockwork cosmos we still need to understand what makes the mechanism tick and that could well require understanding how the world is "in itself" mathematical. Might natural necessitation in configuration space be a form of implication? In the meantime, I shall attempt to describe the cogs in a quantum-mechanical clockwork.

3 The Quantum Multiverse

Thinking about observers in Tegmark's Level I "cosmological" multiverse has been a useful conceptual stepping stone for introducing Unitary Mind and the associated Concrete Sets hypothesis. It opens the way to thinking about different sorts of multiverse. Of course, the cosmological multiverse may well exist, if so it adds a further layer of structure to what follows, but best to set that aside for now.

Charles Sebens considers a *quantum* multiverse which consists of a set of independently-evolving Pilot Wave universes (*op.cit.*: §3). Everettian branching becomes the partitioning of the set of Pilot Wave universes into subsets. What Sebens demonstrates in the subsequent section, with acknowledged limitations, is that the set of independently evolving Pilot Wave universes could be replaced by a set of interacting universes with the interactions mediated by "Newtonian" forces. So, a mixed particle-and-wave ontology is replaced by a particle only ontology; the wave aspect of unitary evolution being generated by particle interactions, like waves generated in a vibrating mass of sand. Similar ideas are to be found in (Hall *et al*. 2014; Boström 2015). Many Interacting Worlds theory, like Pilot Wave theory, only involves objects with definite physical properties, which might be the properties of fields and/or particles but I shall speak in terms of particles for the sake of convenience. The set-theoretic metaphysics can overcome the acknowledged limitations of extant Many Interacting Worlds theories.

For the set-theoretic quantum multiverse the observer's *environmental* universe becomes a set of *elemental* universes which includes all possible particle configurations consistent with observations; in particular, observations of Born rule probabilities. What a wavefunction *is* is a set of *elemental* particle configurations, each in an *elemental* universe. Like the Cheshire cat, the *environmental* free electron teases the observer with its grin; what exists in the observer's spatial environment is a set whose elements are elsewhere, in configuration space.

Consider, for instance, the wavefunction of an *environmental* electron arriving at an *environmental* detector array after having passed through a double-slit apparatus. The absolute square of amplitude of the wavefunction yields the probability that a particular *enviromental* detector will be observed to fire. The reason that's so, according to the set-theoretic metaphysics, is that the *environmental* detector array is a set of *elemental* detector arrays for each of which an *elemental* electron following a trajectory is about to impact an *elemental* detector. And the measure of the subset of *elemental* arrays where a particular *elemental* detector is about to be impacted is the probability that the *environmental* detector, which is the set of those *elemental* detectors, will be observed to fire.

Talk of probabilities here presumes measures on *infinite* sets. Is an *environmental* object which is constituted by a set of *elemental* objects with definite *elemental* particle configurations an infinite set? That would seem to depend on whether spacetime is continuous.

However, even if it's not, and the relevant configurations are finite in number, the possibility of the cosmological multiverse can be invoked until such time as we have evidence that space is finite. That provides a denumerably infinite number of *elemental* objects for every configuration.

On observing the impact of the *environmental* electron on the *environmental* array the observer's *environmental* body partitions into subsets of *elemental* bodies, each subset being the body of an observer observing the firing of a post-impact *environmental* detector. The environment of the pre-impact observer partitions into the environments of the many post-impact observers and the subset measures of those environments relative to the pre-impact environment are the probabilities of the observations of particular post-impact *environmental* detectors firing. The interference pattern for a single *environmental* electron arriving at an *environmental* array is manifest as the distribution of measures of the subsets of *elemental* arrays corresponding to each detection site. When the *environmental* electron passes through the *environmental* slits there are two subsets of *elemental* electrons each of which passes through one *elemental* slit. According to Many Interacting Worlds theory, it's the interactions between the *elemental* electrons in those two subsets which gives rise to the interference pattern.

3.1 Spin, Separability and Entanglement

An *environmental* electron may have indefinite spin but an *elemental* electron cannot, since all *elemental* objects are in definite states. That being so, any *elemental* electron must be either spin-up or spin-down relative to some particular orientation, like a classical spinning object[3]. Given Concrete Sets, an *environmental* z-spin-up electron must have a subset of *elemental* electrons with definite spins relative to the z-axis and all the *elemental* electrons in that subset must be spin-up, otherwise the *environmental* electron wouldn't have definite z-spin. Furthermore, the *environmental* electron must have a subset of *elemental* electrons for every other orientation and each of those subsets must have subsets of spin-up and spin-down *elemental* electrons since the *environmental* electron has indefinite spin relative to those orientations. Necessarily the subset measures of spin-up and spin-down *elemental* electrons for a given orientation must be the probabilities for the *environmental* electron to be measured spin-up or spin-down relative to that orientation.

---

[3] Sebens introduces this idea in the context of his version of Many Interacting Worlds theory (*op.cit.*: §13).

Consider a pair of *environmental* electron in a singlet state. It's a set of *elemental* electrons such that it has two subsets which are *environmental* electrons that have anti-correlated spins. Each of the two *environmental* electrons is a set of *elemental* electrons. Their spins are correlated but that's no reason to suppose that they have any elements in common; they are distinct particles; they're *separable*. The wavefunction of the pair of *environmental* electrons is constituted by the two distinct sets of *elemental* electrons. If one of the *environmental* electrons is measured spin-up relative to some orientation then the second *environmental* electron must be measured spin-down relative to that orientation. If the second *environmental* electron has its spin measured relative to some other orientation then its subset of *elemental* electrons for that orientation must have subsets of spin-up and spin-down *elemental* electrons whose measures correspond to the Bell correlations.

3.2 Locality

The set-theoretic metaphysics involves an *environmental* universe inhabited by observers which is constituted by a set of *elemental* universes where all objects have definite properties. The *elemental* universes are like those of Pilot Wave or Many Interacting Worlds theory, where it's generally acknowledged that there's nonlocality in the form of instantaneous action at a distance. The question now is, does causal nonlocality still apply when observers are alternatively situated in the way I've described? If that's the case, a change in the properties of objects which exist in some *environmental* region A can cause a change in the properties of objects in another *environmental* region B, where A and B are spacelike separated.

The EPR-Bell setup, from the Unitary Mind + Concrete Sets point of view, brings a new dimension to this issue, namely configuration space as an actually existing set of configurations. Objects in observers' environments are set-theoretically "extended" in configuration space.

So, we have Alice in region A and Bob in region B making spin measurements on an electron singlet state created in their common past. In regions A and B there exist parts of the singlet's wavefunction and so subsets of both the *environmental* electrons. Prior to Alice's measurement, both those *environmental* electrons have indefinite spin relative to all orientations, and so each has a subset of *elemental* electrons for each orientation and each of those subsets has spin-up and spin-down subsets of equal measure.

On making her measurement relative to some orientation ô, Alice measures BOTH *environmental* electrons at the same time. The particles are fungible but nonetheless distinct, so Alice's body partitions into two subsets, each of which is a body interacting with one of the two *environmental* electrons. Each of those subsets further partitions into two subsets, one registering ô-spin-up and the other ô-spin-down. So it would seem that there should be *four* downstream observers, two observing ô-spin-up and two observing ô-spin-down. However, Unitary Mind ensures that there are just two downstream observers, Alice$_{UP}$ seeing ô-spin-up and Alice$_{DOWN}$ seeing ô-spin-down. Notice that Alice$_{UP}$'s and Alice$_{DOWN}$'s observations do not determine a particular one of the two *environmental* electrons as being ô-spin-up or ô-spin-down, the *identity* the measured *environmental* particle is *indefinite* relative to Alice$_{UP}$ and Alice$_{DOWN}$.

Knowing her quantum theory and the work of Alain Aspect, Alice will know that Alice$_{UP}$ will eventually see Bob$_{DOWN}$ and Alice$_{DOWN}$ will eventually see Bob$_{UP}$, who will have fissioned from Bob, making his measurement in region B.

So there must indeed have been a change in region B when Alice made her measurement in region A. Bob measures two *environmental* particles which have opposite ô-spins, whereas Alice measures two *environmental* particles which have indefinite spins for all orientations. If Bob had measured the spins of the *environmental* electrons in region B relative to some different orientation the Bell correlations would still apply because of the *elemental* structure of *environmental* electrons with definite ô-spin. A definite ô-spin environmental electron has a subset of *elemental* electrons which are all ô-spin-up or all ô-spin-down and for any subset of *elemental* electrons which have definite spins for some other orientation ô$^*$ the measures of ô$^*$-spin-up and ô$^*$-spin-down *elemental* electrons are determined according the angle between ô and ô$^*$. Those measures are the probabilities of measuring ô$^*$-spin-up and ô$^*$-spin-down for a definite ô-spin *environmental* electron.

So the set-theoretic metaphysics is, after all, nonlocal, *pace* (Tappenden 2023: §4.3). But it's not, after all, nonlocality which raises difficulties for conventional hidden-variable theories; it's the lack of superposition. For hidden-variable theories where observers inhabit hidden-variable universes, Alice's measurement in region A causes a change in a *definite* state in region B, but for the set-theoretic metaphysics the change is in an *indefinite* state. As a result, when Bob makes his measurement in region B his body evolves into an indefinite macroscopic state in Alice$_{UP}$'s and Alice$_{DOWN}$'s absolute elsewheres, an indefinite *environmental* object relative to each of them. Likewise, post-measurement Alice's body is an indefinite *environmental* macroscopic object relative to Bob$_{UP}$ and Bob$_{DOWN}$. The set-theoretic metaphysics allows for indefinite *environmental* macroscopic objects in observers' unobserved, or unobservable, environments.

David Deutsch once wrote:

> pilot-wave theories are parallel-universes theories in a state of chronic denial.
>
> (Deutsch 1996: 226)

Those words close a paragraph where he describes Pilot Wave theory as only filling one of the "grooves" in an "immensely complicated multi-dimensional wavefunction" and claims that "the 'unoccupied grooves' must be physically real". A retort could be that parallel-universes theories have been hidden-variable theories in denial. For Deutsch's unoccupied grooves to be physically real they need to be occupied. The set-theoretic metaphysics fills *all* the grooves by making the *environmental* wavefunction a set of *elemental* objects.

4  Conclusion and Consequences

The metaphysics which I've described brings a new dimension to physics, what Michael Lockwood gestured towards as a *Pickwickian* dimension, attributing the idea to Deutsch (Lockwood 1989: 232). I've argued that it can be thought of as configuration space actualised, a partially-ordered dimension. Physical objects in an observer's environment thereby become "extended" in configuration space in the sense that they are sets of objects with different configurations. An *environmental* free electron is a set of *elemental* electrons on different trajectories. A macroscopic *environmental* object is a set of macroscopic *elemental* objects, each constituted by a different configuration of particles. The body of an observer is a set of doppelgängers which partitions into cognitively distinct subsets in measurement contexts, so the pre-measurement observer bears the relation *will be* to each of the post-measurement observers making different observations.

Odd as it may seem, this changes nothing in everyday life. Different futures have probabilities, just as before, the only difference is that they're not fictional possibilities, they're actualities. Actors will still tend to strive to make the futures they desire more probable. The notion of free will remains as challenging as it was in classical, deterministic-but-not-probabilistic physics.

However, there's an elephant in the room of Many Worlds theory which can't be ignored, going by the names of *quantum suicide* or *quantum Russian roulette*. Theorists have blown hot and cold on the subject over the last few decades (Tegmark 1997, Carroll *op.cit.*: 209). For a quantum Russian roulette setup, the set-theoretic metaphysics has it that there's a survival branch with an objective probability of 1/6 and the person pulling the trigger can be sure that they will be a person who has survived. The best argument to the contrary, so far as I know, is in (Papineau 2003). A response is to be found in (Tappenden 2004).

David Lewis concluded:

> You who bid good riddance to collapse laws, you quantum cosmologists, you enthusiasts of quantum computing, should shake in your shoes. Everett's idea is elegant, but heaven forefend that it should be true!
>
> (Lewis 2004: 21)

For quantum statistical mechanics with actualised configuration space every erstwhile "possible" physical future has some probability, be it ever so small. There are always survival branches, even when you hit the deck after falling from an aircraft without a parachute, or when you're lingering in a hospice with "terminal" cancer. "Aye, there's the rub". Does some sort of eternal *terminal limbo* await us all,rather than oblivion? Coping with that prospect is a challenge we may have to face.